\documentclass[nonacm, sigconf]{acmart}

\usepackage{tikz-cd}
\usepackage{lipsum}
\usepackage{wasysym}
\usepackage{tcolorbox}
\usepackage{hyperref}

\AtBeginDocument{%
  \providecommand\BibTeX{{%
    \normalfont B\kern-0.5em{\scshape i\kern-0.25em b}\kern-0.8em\TeX}}}

\setcopyright{acmcopyright}
\copyrightyear{2018}
\acmYear{2018}
\acmDOI{XXXXXXX.XXXXXXX}

\acmConference[Conference acronym 'XX]{Make sure to enter the correct
  conference title from your rights confirmation emai}{June 03--05,
  2018}{Woodstock, NY}
\acmPrice{15.00}
\acmISBN{978-1-4503-XXXX-X/18/06}

\begin{document}

%%
%% The "title" command has an optional parameter,
%% allowing the author to define a "short title" to be used in page headers.
\title{Os efeitos da arquitetura dirigida a eventos na modularidade de software: Um estudo exploratório}

%%
%% The "author" command and its associated commands are used to define
%% the authors and their affiliations.
%% Of note is the shared affiliation of the first two authors, and the
%% "authornote" and "authornotemark" commands
%% used to denote shared contribution to the research.
% \author{Ben Trovato}
% \authornote{Both authors contributed equally to this research.}
% \email{trovato@corporation.com}
% \orcid{1234-5678-9012}
% \author{G.K.M. Tobin}
% \authornotemark[1]
% \email{webmaster@marysville-ohio.com}
% \affiliation{%
%   \institution{Institute for Clarity in Documentation}
%   \streetaddress{P.O. Box 1212}
%   \city{Dublin}
%   \state{Ohio}
%   \country{USA}
%   \postcode{43017-6221}
% }

\author{Luan Lazzari}
\affiliation{%
  \institution{Universidade do Vale do Rio do Sinos (Unisinos)}
  \streetaddress{Av. Unisinos, 950 - Cristo Rei}
  \city{São Leopoldo, Rio Grande do Sul}
  \country{Brasil}}
\email{luanlazzari@hotmail.com}

\author{Kleinner Farias}
\affiliation{%
  \institution{PPGCA, Universidade do Vale do Rio do Sinos (Unisinos)}
  \streetaddress{Av. Unisinos, 950 - Cristo Rei}
  \city{São Leopoldo, Rio Grande do Sul}
  \country{Brasil}}
\email{kleinnerfarias@unisinos.br}

%%
%% By default, the full list of authors will be used in the page
%% headers. Often, this list is too long, and will overlap
%% other information printed in the page headers. This command allows
%% the author to define a more concise list
%% of authors' names for this purpose.
\renewcommand{\shortauthors}{Lazzari and Farias}

%%
%% The abstract is a short summary of the work to be presented in the
%% article.
\begin{abstract}
Este trabalho apresenta um estudo exploratório sobre os efeitos da arquitetura dirigida a eventos na modularização de \textit{software}. Ele é um estudo inicial do qual se busca compreender os efeitos da adoção de arquitetura orientada a eventos na separação de interesses, complexidade, acoplamento, coesão e tamanho, em comparação com o estilo de arquitetura REST. Uma aplicação foi desenvolvida usando a arquitetura dirigida a eventos e a arquitetura tradicional REST através de cinco cenários de evolução. Em cada cenário, uma funcionalidade foi adicionada. As versões geradas foram comparadas usando 10 métricas. Até onde sabemos, os resultados relatados são os primeiros a descrever os benefícios da arquitetura dirigida a eventos, em termos de modularidade de \textit{software} em cenários reais de evolução. Nesse caso, o estudo pode ser visto como o primeiro passo em uma agenda mais ambiciosa para avaliar os benefícios da arquitetura empiricamente orientada a eventos.
\end{abstract}

%%
%% The code below is generated by the tool at http://dl.acm.org/ccs.cfm.
%% Please copy and paste the code instead of the example below.
%%
% \begin{CCSXML}
% <ccs2012>
%  <concept>
%   <concept_id>10010520.10010553.10010562</concept_id>
%   <concept_desc>Computer systems organization~Embedded systems</concept_desc>
%   <concept_significance>500</concept_significance>
%  </concept>
%  <concept>
%   <concept_id>10010520.10010575.10010755</concept_id>
%   <concept_desc>Computer systems organization~Redundancy</concept_desc>
%   <concept_significance>300</concept_significance>
%  </concept>
%  <concept>
%   <concept_id>10010520.10010553.10010554</concept_id>
%   <concept_desc>Computer systems organization~Robotics</concept_desc>
%   <concept_significance>100</concept_significance>
%  </concept>
%  <concept>
%   <concept_id>10003033.10003083.10003095</concept_id>
%   <concept_desc>Networks~Network reliability</concept_desc>
%   <concept_significance>100</concept_significance>
%  </concept>
% </ccs2012>
% \end{CCSXML}

% \ccsdesc[500]{Computer systems organization~Embedded systems}
% \ccsdesc[300]{Computer systems organization~Redundancy}
% \ccsdesc{Computer systems organization~Robotics}
% \ccsdesc[100]{Networks~Network reliability}

%%
%% Keywords. The author(s) should pick words that accurately describe
%% the work being presented. Separate the keywords with commas.

%% A "teaser" image appears between the author and affiliation
%% information and the body of the document, and typically spans the
%% page.
% \begin{teaserfigure}
%   \includegraphics[width=\textwidth]{sampleteaser}
%   \caption{Seattle Mariners at Spring Training, 2010.}
%   \Description{Enjoying the baseball game from the third-base
%   seats. Ichiro Suzuki preparing to bat.}
%   \label{fig:teaser}
% \end{teaserfigure}

%%
%% This command processes the author and affiliation and title
%% information and builds the first part of the formatted document.
\maketitle

\section{Introdução}

A arquitetura dirigida a eventos é uma solução promissora para o desenvolvimento de sistemas distribuídos que entrega modularização, escalabilidade e concorrência \cite{Ludger}. O KAFKA\footnote{KAFKA: https://www.confluent.io/} seria um exemplo de tecnologia que suporta arquitetura dirigida a eventos, propondo uma série de componentes que se comunicam através de eventos \cite{Simon}. Logo, o processamento procura compor serviços, não por meio de cadeias de comandos e consultas, mas sim pelo fluxo de eventos \cite{stopford2018designing}. De tal forma que cada componente pode executar a sua tarefa independente, pois são acionados por gatilhos, que por sua vez são eventos \cite{Simon}. Ao ponto em que há a separação entre computação e comunicação, tornando fácil a integração entre componentes heterogêneos em sistemas complexos que são fáceis de evoluir e escalar \cite{Ludger}, como por exemplo, ambientes inteligentes \cite{schipor2019euphoria}.  

Já em modelos de sistemas baseados em requisição/resposta, os dados são obtidos em diferentes fontes via requisições, por exemplo, HTTP, podendo gerar possíveis congestionamentos \cite{Ludger}. A literatura atual~\cite{stopford2018designing} aponta que a arquitetura dirigida a eventos promove o fraco acoplamento, essencial para a modularização de software, porém pode aumentar a complexidade do projeto e o entendimento do sistema \cite{Ludger}. Entre as arquiteturas tradicionais para implementação de sistemas orientados a serviços, destaca-se a arquitetura REST~\cite{fielding2000architectural}. 

Os proponentes da arquitetura dirigida a eventos advogam que projetar aplicações fortemente baseada em eventos favorece à modularização das funcionalidades, bem como facilita atividades de manutenção e evolução. Neste sentido, projetar software adotando uma arquitetura dirigida a eventos pode implicar em uma forma mais sistemática de promover uma melhor modularização de software. Atualmente, conjectura-se que o uso de tecnologias como o KAFKA gerará aplicações com uma melhor separação de interesses, melhor acoplamento e coesão, menor complexidade e tamanho. No entanto, há poucas evidências para confirmar se essa expectativa se confirma ou não. Hoje, a literatura ainda carece de estudo exploratórios que investiguem os efeitos de uma arquitetura dirigida a eventos em aspectos de modularidade de software. Além disso, não se tem conhecimento se esses efeitos são melhores ou piores que os causados por arquiteturas tradicionais e amplamente utilizadas, tal com o estilo arquitetura REST. Consequentemente, os desenvolvedores acabam adotando arquitetura dirigida a eventos sem nenhuma evidência empírica sobre seus efeitos na modularidade de software. Alguns estudos na literatura apontam para alguns benefícios da adoração de arquitetura dirigida a eventos. Laigner et al.~\cite{laigner2020monolithic} reporta um estudo empírico no qual se constatou que a adoção da arquitetura dirigida a eventos melhorou a manutenção e o isolamento de falhas em um sistema que foi refatorado após passar anos evoluindo, dando origem a um código grande, complexo e obsoleto, exigindo um processo de manutenção caro. Urdangarin et al.~\cite{urdangarin2021mon4aware} destaca a importância da decomposição de aplicações monolíticas, principalmente para reduzir o esforço de manutenção de software~\cite{eduardo2020software}.

Este artigo, portanto, apresenta um estudo empírico exploratório sobre os efeitos de arquitetura dirigida a eventos na modularidade de software. Trata-se de um estudo inicial através do qual  busca-se compreender os efeitos da adoção arquitetura dirigida a eventos na separação de interesses, complexidade, acoplamento, coesão e tamanho, em comparação com o estilo arquitetura REST. Investiga-se uma faceta particular da arquitetura dirigida a eventos em relação aos benefícios na evolução de uma aplicação alvo real através da adição de novas funcionalidades. Uma aplicação foi desenvolvida utilizando uma arquitetura dirigida a eventos e uma arquitetura tradicional REST através de 5 cenários de evolução. Em cada cenário, uma funcionalidade foi adicionada. As versões geradas foram comparadas usando 10 métricas. Os resultados reportados são os primeiros a relatar as vantagens potenciais da arquitetura dirigida a eventos, em termos de modularidade de software em cenários reais de evolução. Neste sentido, este artigo pode ser visto como um primeiro passo em uma agenda mais ambiciosa para avaliar os benefícios da arquitetura dirigida a eventos empiricamente.

O estudo está dividido conforme a seguinte estrutura: a Seção~\ref{sec:background} conterá o referencial teórico, com os principais conceitos para entendimento do estudo proposto; a Seção~\ref{sec:related_works} abordará os trabalhos relacionados, explorando o processo de seleção utilizado e também realizando um comparativo destes com o presente; a Seção~\ref{sec:metodologia} abordará a descrição da metodologia para desenvolvimento do estudo; a Seção~\ref{sec:resultados} traz os resultados obtidos; e, por fim, a Seção~\ref{sec:conclusion} traça algumas conclusões e trabalhos futuros.

\section{Referencial teórico}
\label{sec:background}

Esta seção aborda os conceitos teóricos usados durante a construção e desenvolvimento do estudo.

\subsection{Modularização de interesses}
\label{subsec:modularization}

Modularização é considerado um conceito essencial no \textit{design} de \textit{software} moderno \cite{SANTANNA}. Definido pelo IEEE como o grau em que um programa de sistema é composto de componentes discretos, de forma que uma mudança em um componente tenha impacto mínimo em outros \cite{SANTANNA}. Enquanto um interesse é qualquer propriedade importante ou área de interesse de um sistema que se deseja tratar de forma modular \cite{SANTANNA}. Portanto, um interesse de \textit{software} pode ser uma \textit{feature}, regra de negócio, requisito não funcional, um padrão de arquitetura ou padrões de projeto \cite{SANTANNA}. 

\textit{Software} com alto grau de modularização traz vários benefícios como compreensão, extensibilidade, adaptabilidade, reuso, entre outros \cite{parnas}. Por consequência a modularização pode ser aplicada em várias etapas do \textit{design}, variando da especificação da arquitetura ao detalhamento do \textit{design} e níveis de abstração de código \cite{SANTANNA}. O principal objetivo da arquitetura de \textit{software} é definir em quais componentes o sistema deve consistir, como esses componentes irão se comunicar uns com os outros e como eles devem ser implantados para cumprir os requisitos \cite{Falatiuk}. Portanto, a arquitetura de \textit{software} desempenha um papel importante na formulação e desenvolvimento de \textit{software} \cite{Gardazi}.

Todavia a decomposição arquitetural ainda é um grande gargalo para o processo de \textit{design} de \textit{software} segundo \cite{SANTANNA}, principalmente pela necessidade da modularização simultânea de uma série de questões de amplo escopo. Consequentemente, a inadequada modularização de interesses pode gerar complexidade no \textit{design} de \textit{software} \cite{SANTANNA}. A avaliação de diferentes arquiteturas requer técnicas para mensurar quantitativamente \cite{SANTANNA}. De modo que as métricas de \textit{software} são poderosos indicadores de modularização no \textit{design} de \textit{software} \cite{SANTANNA}. Logo, a comunidade tem utilizado consistentemente noções de acoplamento de módulo, coesão e tamanho de interfaces para mensurar a modularização \cite{SANTANNA}. Por isso, neste estudo foi definido um conjunto de métricas abordadas na Seção~\ref{sec:metodologia-metricas}.

\subsection{Arquitetura dirigida a eventos}
\label{subsec:arq-eventos}

No modo baseado em eventos, os componentes apenas publicam dados, sem conhecer os demais componentes, muitos menos quem irá consumir e reagir aos dados, promovendo a separação da computação e publicação de eventos de qualquer processamento subsequente \cite{Ludger}. Além disso, sua comunicação é assíncrona, no modelo produtor/consumidor, ambos são independentes um do outro \cite{Falatiuk}. Por consequência, promovendo o fraco acoplamento entre os componentes, motivo pelo qual a arquitetura dirigida a eventos tornou-se predominante em aplicações distribuídas em grande escala \cite{Ludger}.

O sistema de mensagens nos permite construir serviços fracamente acoplados, pois ele move os dados puros a um local altamente acoplado (o produtor) e coloca em um local fracamente acoplado (o consumidor) \cite{stopford2018designing}. Portanto, quaisquer operações que precisem ser executadas nesses dados não são feitas no produtor, mas em cada consumidor \cite{stopford2018designing}. Isto é, serviços podem ser facilmente adicionados ao sistema, no modo \textit{plug and play} (plugável), em que ele se conecta aos fluxos de eventos e executando quando seus critérios são atendidos \cite{stopford2018designing}.

Não só promove o baixo acoplamento mas também o \textit{Kafka Stream} é capaz de armazenar os eventos e dados, não necessitando um banco de dados externo, e, mantendo os eventos ``próximos'' dos serviços \cite{stopford2018designing}. Além disso, todos os eventos são armazenados na ordem em que chegaram, permitindo que os eventos sejam reproduzidos em ordem \cite{stopford2018designing}. Por isso o desempenho de aplicações baseadas em eventos também se mostra melhor, garantindo estabilidade e alto desempenho para grandes quantidades de eventos.

\subsection{Arquitetura \textit{REST}}
\label{subsec:arq-REST}

Dentre as arquiteturas tradicionais, destaca-se a \textit{REST} no modelo requisição/resposta síncrono. Embora, também seja possível operar no modelo assíncrono seu uso é menos difundido. No modelo síncrono o cliente realiza uma requisição e aguarda pela resposta, enquanto está sendo processado pelo serviço responsável. De forma que é amplamente utilizada em aplicações distribuídas em rede \cite{Zhou}. Promovendo, portanto, o fraco acoplamento entre seus interesses, um dos motivos para sua popularidade \cite{Zhou}. Tanto que dentre os modelos arquitetônicos mais utilizados está o modelo MVC (Model–View–Controller), amplamente difundido \cite{Spaccapietra}. Pois, seu principal conceito é a separação em camadas, separando a persistência de dados, interface do usuário e controle da aplicação \cite{Ping}. Tal separação em camadas tem como objetivo promover a separação de interesses e consequentemente promover a modularização. 

Diferentemente de arquiteturas dirigidas a eventos, adicionar novos serviços na arquitetura \textit{REST}, em geral, implica em introduzir um novo método e realizar chamadas pelos serviços que precisam dele \cite{stopford2018designing}. Entretanto, a arquitetura \textit{REST} possui algumas vantagens consideráveis como a simplicidade em implementar, dados (ou estado) reside em apenas um lugar e controle centralizado \cite{stopford2018designing}. De forma que neste estudo o principal ponto de comparação entre a arquitetura \textit{REST} e a orientada a eventos será a modularização. A fim de mensurar a modularização, alguns conceitos de qualidade de software e métricas serão abordados para suporte da análise quantitativa.

\section{Trabalhos Relacionados}
\label{sec:related_works}

A pesquisa pelos trabalhos relacionados foi realizada em repositórios digitais como \textit{Google Scholar} e \textit{Scopus (Elsevier)}. Grande parte dos trabalhos selecionados foram resultados de pesquisas pelo termo ``event-driven architecture  AND  microservice''.

\subsection{Análise dos trabalhos relacionados}
\label{subsec:analysis_of_RW}

\textbf{Figueiredo~\textit{et al.}~\cite{Figueiredo}.} Apresenta um estudo quantitativo sobre duas \textit{Software Product Lines (SPLs)}, a fim de avaliar várias facetas de estabilidade de \textit{design}, considerando métricas como SoC (\textit{Separation of Concerns}), acoplamento, e coesão. As SPLs foram implementadas usando programação orientada a aspectos (AO) e orientada a objetos (OO). As versões AO e OO SPLs foram comparadas, buscando entender os benefícios de AO em questões de qualidade de software. O artigo reporta os benefícios de uma arquitetura orientada a aspectos de SPLs. 

\textbf{Garcia \textit{et al.}~\cite{Garcia}.} Apresentam um estudo quantitativo, comparando implementações em Java e AspectJ dos 23 padrões de projeto da \textit{Gang-of-Four (GoF)}. Para tal foi utilizada a programação orientada a objetos (OO) e programação orientada a aspectos (AOP). A fim de comparação, as métricas utilizadas foram o acoplamento dos objetos e o SoC, mas também a coesão e tamanho. Considerando as características das implementações em cada padrão. Após cada alteração eram coletadas as métricas, comparando sempre com a versão anterior, antes das alterações. Por fim, reporta em quais pontos AOP se destacou positivamente e negativamente em comparação ao OO.

\textbf{Fiege \textit{et al.}~\cite{Ludger}.} Apresenta um estudo qualitativo sobre o \textit{design} modular e a implementação de um sistema de eventos, capaz de suportar escopos e mapeamentos de eventos. Dentre os conceitos, são especificados o papel do produtor e consumidor, formas de comunicação entre eles e gatilhos entre eventos. Bem como os benefícios, os quais pode-se citar a modularização do sistema, baixo acoplamento, abstração e ocultamento de informações. Além dos componentes que compõem uma arquitetura dirigida a eventos, como o \textit{subscribe/unsubscribe} dos eventos, necessário para garantir a distribuição de mensagens. Bem como o ponto central responsável por gerenciar parte do sistema, como passar um gatilho a um ou mais eventos.

\textbf{Falatiuk \textit{et al.}~\cite{Falatiuk}.} Apresentam um estudo qualitativo, descrevendo os principais conceitos arquiteturais e seleção de tecnologias para implementação de um sistema de gerenciamento de documentos, o \textit{e-archive}. Para isso, as arquiteturas escolhidas foram a dirigida a eventos e microsserviços, de acordo com os requisitos levantados. Como vantagens encontradas, estão a escalabilidade horizontal, modularização, perda de acoplamento entre componentes, facilidade de modificações e manipulação de grande quantidade de dados. Contudo, requer maior conhecimento de padrões de arquitetura na nuvem e cultura \textit{DevOps}, para garantir o escalonamento e o devido monitoramento. Assim, o \textit{design} se mostra bastante custoso em seu início, mas oferece futuras manutenções, modificações e atualizações mais baratos  à medida que o sistema evoluir.

\textbf{Alaasam \textit{et al.}~\cite{Alaasam}.} Propõe um estudo de caso sobre a viabilidade de uso do \textit{Apache  Kafka Stream  API (Kafka  stream  DSL)} no desenvolvimento do \textit{Digital Twin (DT)}. Um sistema de processamento de fluxos de dados em tempo real, capaz de monitorar, controlar e prever estados a partir de dados, coletados de diversos sensores. Nele foi realizado um estudo paramétrico de latência e tempo de resposta, levando em consideração tolerância a falhas, escalabilidade e facilidade de implementação. Como conclusão, o \textit{Kafka} se mostrou adequado ao sistema proposto, fornecendo um bom gerenciamento de estados e latência aceitável. Entretanto, percebeu-se uma perda na eficiência, enquanto há muito tráfego de dados entre os tópicos intermediários. 

\textbf{Schipor \textit{et al.}~\cite{schipor2019euphoria}.} Introduz o \textit{Euphoria}, um nova arquitetura de \textit{software} orientado a eventos, voltado aos ambientes inteligentes. Compostos por uma enorme gama de dispositivos heterogêneos, cada um com seu sistema operacional, protocolos de comunicação, forma de interação entre outros. Tais ambientes possuem alguns critérios de \textit{design} como a modularidade, escalabilidade e assincronicidade para produzir, processar e transmitir mensagens e eventos. Portanto, o \textit{Euphoria} foi projetado adotando várias técnicas e propriedades de qualidade seguindo aos padrões do \textit{(SQuaRE) ISO/IEC 25000}. Além disso, foi conduzido uma avaliação técnica sobre as capacidades do \textit{Euphoria}, foram quantificados tamanho das mensagens, diferentes dispositivos e complexidade do ambiente (quantidade de dispositivos). Seu resultado foi satisfatório, alcançando um baixo tempo de resposta mesmo em um ambiente composto por um número grande de produtores e consumidores.

\textbf{Laigner \textit{et al.}~\cite{laigner2020monolithic}.} Realizou a troca de um sistema de \textit{Big Data} legado (BDS) para uma arquitetura orientada a eventos baseada em microsserviços. Tal BDS está localizado no Instituto Tecgraf da PUC-Rio, que fornece soluções para parceiros industriais. Uma das soluções desenvolvidas para um cliente do setor de Óleo e Gás em 2008, diz respeito a um BDS que monitora objetos em movimento e detecta de forma proativa eventos que geram riscos à operação, como desvios de rota de veículos. Motivados pelo difícil processo de manutenção do sistema e o advento de um novo parceiro, desenhou-se a reescrita completa do BDS legado para então tecnologias atuais de \textit{Big Data}. Como conclusão, observou-se o suporte a microsserviços para manutenção mais fácil e isolamento de falhas como benefícios. No entanto, o fluxo de dados complexo gerado pelo número de microsserviços, bem como a miríade de tecnologias, como desvantagens.

\subsection{Análise comparativa dos trabalhos relacionados}
\label{subsec:comparative-related-works}

\textbf{Critério de Comparação.} Foram definidos cinco Critérios de Comparação (CC) para realizar a comparação de similaridades e diferenças entre o trabalho proposto e os artigos selecionados. Sua comparação é crucial para auxiliar na identificação de oportunidades de pesquisa utilizando critérios objetivos, ao invés de subjetivos. Os critérios são descritos a seguir:

\begin{itemize}
    \item \textbf{Estudo Empírico (CC1):} pode ser entendida como aquela em que é necessária comprovação prática de algo, especialmente por meio de experimentos ou observação de determinado contexto para coleta de dados em campo;
    \item \textbf{Análise de Modularização (CC2):} compreende a capacidade de decomposição da aplicação em programas menores com interfaces padronizadas, oferecendo maior capacidade de gerenciamento no desenvolvimento da aplicação;
    \item \textbf{Arquitetura dirigida a eventos (CC3):} composta por diversos componentes onde a comunicação entre produtores e consumidores ocorre por meio de eventos;
    \item \textbf{Arquitetura de microsserviço (CC4):} cada módulo da aplicação é totalmente independente de outros (\textit{standalone}), cada contém uma suíte de serviços autônomos, comunicando-se por meio de uma API;
    \item \textbf{Contexto de aplicação (CC5):} define se o estudo aborda a implementação de uma aplicação real.
\end{itemize}

\textbf{Oportunidades de pesquisa.} A Tabela~\ref{tab:tabela-comparativa} apresenta a comparação dos estudos selecionados, evidenciando o não atendimento de todos os critérios. A seguir, as oportunidades observadas a partir das comparações: (1) apenas o trabalho proposto atende todos os critérios de comparação definidos; (2) nenhum estudo empírico referente à arquitetura dirigida a eventos; e (3) assim como nenhum, sobre modularização de \textit{software} em arquitetura dirigida a eventos e microsserviços. Portanto, a seguinte oportunidade de pesquisa foi identificada: análise sobre os efeitos da arquitetura dirigida a eventos na modularização de \textit{software}. À medida que será explorada nas próximas seções.

\begin{table}[!ht]
    \footnotesize
    \centering
    \caption{Análise comparativa dos Trabalhos Relacionados selecionados}
    \begin{tabular}{|l|c|c|c|c|c|c|}
        \hline
        %\multicolumn{1}{|c}{\multirow{2}{*}{\textbf{Trabalho Relacionado}}}
        \multicolumn{1}{|c}{\textbf{Trabalho Relacionado}}
        & \multicolumn{5}{|c|}{\textbf{Critério de Comparação}} \\\cline{2-6}
            & \textbf{CC1}   & \textbf{CC2}  & \textbf{CC3}  & \textbf{CC4} & \textbf{CC5}  \\ \hline
        Trabalho Proposto    & $\CIRCLE$      & $\CIRCLE$     & $\CIRCLE$     & $\CIRCLE$    & $\CIRCLE$     \\ \hline
        Figueiredo~\textit{et al.}~\cite{Figueiredo}   & $\CIRCLE$      & $\CIRCLE$     & $\Circle$    & $\Circle$     & $\CIRCLE$    \\ \hline
        Garcia \textit{et al.}~\cite{Garcia}            & $\CIRCLE$      & $\CIRCLE$     & $\Circle$    & $\Circle$     & $\CIRCLE$    \\ \hline
        Fiege \textit{et al.}~\cite{Ludger}            & $\LEFTcircle$  & $\CIRCLE$     & $\CIRCLE$    & $\Circle$     & $\Circle$    \\ \hline
        Falatiuk \textit{et al.}~\cite{Falatiuk}        & $\Circle$      & $\LEFTcircle$ & $\CIRCLE$    & $\CIRCLE$     & $\Circle$    \\ \hline
        Alaasam \textit{et al.}~\cite{Alaasam}          & $\LEFTcircle$  & $\Circle$     & $\CIRCLE$    & $\CIRCLE$     & $\CIRCLE$    \\ \hline
        Schipor \textit{et al.}~\cite{schipor2019euphoria}    & $\CIRCLE$  & $\Circle$    & $\CIRCLE$    & $\Circle$     & $\CIRCLE$    \\ \hline
        Laigner \textit{et al.}~\cite{laigner2020monolithic} & $\CIRCLE$  & $\Circle$    & $\CIRCLE$    & $\CIRCLE$     & $\CIRCLE$    \\ \hline
    \end{tabular}
    \begin{tabular}{ccc}
        $\CIRCLE$ Atende Completamente & $\LEFTcircle$ Atende Parcialmente & $\Circle$ Não Atende
    \end{tabular}
    \label{tab:tabela-comparativa}
%    \fonte{Elaborado pelo autor.}
\end{table}

\section{Metodologia}
\label{sec:metodologia}

Esta seção descreve a metodologia seguida para executar o estudo empírico. A Seção~\ref{subsec:metodologia-objetivo} apresenta o objetivo do estudo e a questão de pesquisa investigada. A Seção~\ref{sec:metodologia-metricas} descreve as métricas utilizadas. A Seção~\ref{sec:metodologia-processo} detalha o processo experimental seguido, descrevendo as fases do estudo e as atividades executadas. A Seção~\ref{sec:metodologia-procedimento} traz os procedimentos de análise dos dados. A Seção~\ref{sec:metodologia-aplicacao} descreve a aplicação alvo escolhida, detalhando suas funcionalidades e características. Por fim, a Seção~\ref{subsec:cenarios-de-mudanca} detalha os cenários de evolução da aplicação alvo. A metodologia deste trabalho é inspirada a partir de estudos empíricos anteriores previamente publicados~\cite{farias2014effects,farias2016empirical,farias2015evaluating,farias2010assessing,oliveira2007guidance,d2020effects,farias2012evaluating}.

\subsection{Objetivo e Questão de Pesquisa}
\label{subsec:metodologia-objetivo}

O objetivo deste estudo é essencialmente avaliar os efeitos da arquitetura dirigida a eventos na modularidade de software. Em particular, busca-se investigar os efeitos sobre cinco diferentes variáveis envolvidas com modularidade~\cite{Figueiredo}: separação de interesses, acoplamento, complexidade, coesão e tamanho. Esses efeitos são investigados no contexto de cenários reais de evolução (Seção~\ref{subsec:cenarios-de-mudanca}) de uma aplicação alvo (Seção~\ref{sec:metodologia-aplicacao}), a qual foi implementada usando a arquitetura dirigida a eventos e usando a arquitetura REST. As duas implementações geradas foram necessárias para viabilizar a comparação. Portanto, o objetivo deste estudo é estabelecido com base no modelo GQM~\cite{basili1992software} da seguinte forma:

\begin{center}
    \textbf{Analisar} estilos arquiteturais \\
    \textbf{com o propósito de} investigar seus efeitos \\
    \textbf{em relação à} modularidade de software\\
    \textbf{através da perspectiva de} desenvolvedores \\
    \textbf{no contexto de} evolução de software.
\end{center}

Em particular, este artigo tenta revelar os efeitos da arquitetura dirigida a eventos na modularidade de software durante a evolução de software. Neste sentido, uma questão de pesquisa (QP) foi formulada:

\begin{itemize}
    \item \textbf{QP:} Arquitetura dirigida a eventos promove uma maior modularização quando comparada ao estilo arquitetura REST?
\end{itemize}

Parnas~\cite{parnas} aponta que, caso a modularização de uma aplicação seja alta, alguns benefícios serão obtidos como, por exemplo, maior facilidade de alteração, maior poder de adaptação e compreensão de código. Além disso, a modularização pode proporcionar o desenvolvimento de cada módulo independentemente, permitindo o desenvolvimento paralelo, redução do tempo de desenvolvimento e o melhor gerenciamento do impacto de alterações~\cite{almentero}. Parnas~\cite{parnas} reforça que um módulo pode ser definido como um conjunto de decisões de projeto independente de outros módulos e a interação entre os módulos deve ser totalmente por meio de suas \textit{interfaces} \cite{parnas} --- promovendo, assim, a separação de interesses e delegando funções isoladas a cada módulo. Portanto, a separação precisa dos interesses da aplicação leva a modularização, que pode tornar a aplicação customizável, permitindo seu uso em diferentes contextos.  Além disso, a modularização permite que o desenvolvedor foque em um módulo por vez, facilitando o entendimento, para depois combinar todos e entender a aplicação pelo todo \cite{almentero}. A arquitetura dirigida a eventos~\cite{stopford2018designing, schipor2019euphoria} tenta contemplar tais benefícios citados, destacando a importância da execução de uma estudo empírico para verificar os benefícios deste novo estilo arquitetural.

\subsection{Métricas}
\label{sec:metodologia-metricas}

\begin{table*}[!ht]
    \footnotesize
    \centering
    \caption{Conjunto de métricas utilizadas no estudo (fonte \cite{Garcia}).}
    \begin{tabular}{|c|p{4cm}|p{9cm}|} \hline
         \textbf{Variável} & \textbf{Métrica} & \textbf{Definição} \\ \hline

         \multicolumn{1}{|c|}{\centering Separação de interesses (SoC)}
         & \textit{Concern Diffusion over components (CDC)} & Conta o número de classes cujo objetivo principal é contribuir para a implementação do cenário e o número de classes que os acessam. \\ \cline{2-3}
         & \textit{Concern Diffusion over operations (CDO)} & Conta o número de métodos cujo objetivo é contribuir para a implementação do cenário e o número de  métodos que os acessam. \\ \cline{2-3}
         & \textit{Concern Diffusion over LOC (CDLOC)} & Conta o número de pontos de transição para implementação do cenário nas linhas de código. Os pontos de transição são pontos no código onde há uma "mudança de preocupação". \\ \hline
         
         \multicolumn{1}{|c|}{\centering Acoplamento}
         & \textit{Coupling between components (Dep\_Out)} & Número de dependências em que o módulo é cliente. \\ \cline{2-3}
         & \textit{Coupling between components (Dep\_In)} & Número de dependências em que o módulo é fornecedor. \\ \hline
         
         \multicolumn{1}{|c|}{\centering Coesão}
         & \centering Relational cohesion (H)  & Mensura o número médio de relacionamentos internos por classe/interface. É calculada como a razão de R + 1 para o número de classes e interfaces por pacote. \\ \hline
         
         \multicolumn{1}{|c|}{\centering Complexidade}
         & Number of relationships (R) & Mensura o número de relacionamentos entre classes e interfaces por pacote. \\ \hline
         
         \multicolumn{1}{|c|}{\centering Tamanho}
         & \textit{Lines of code (LOC)} & Conta as linhas de código cujo objetivo é contribuir para a implementação do cenário. \\ \cline{2-3}
         & \textit{Number of attributes (NumAttr)} & Conta o número de atributos cujo objetivo é contribuir para a implementação do cenário. \\ \cline{2-3}
         & \textit{Weighted operations per component (NumOps)} & Conta o número de operações cujo objetivo é contribuir para a implementação do cenário. \\ \hline
        
    \end{tabular}
    \label{tab:tabela-metricas}
\end{table*}

Tabela~\ref{tab:tabela-metricas} apresenta as métricas utilizadas para quantificar cinco variáveis de modularidade, incluindo separação de interesses (SoC), acoplamento, coesão, complexidade e tamanho. Tais métricas foram utilizadas, pois estudos empíricos anteriores~\cite{Figueiredo, Garcia} já mostraram a validade delas em investigações sobre modularidade de software.

\textbf{Separação de interesses.} Neste estudo, esta métrica busca medir o grau de modularização das funcionalidades implementadas usando a arquitetura dirigida a eventos e o estilo arquitetural REST. A SoC utilizará três métricas: (i) componentes (ou classes) baseado no \textit{Concern Diffusion over components (CDC)}, (ii) operações (ou funções) baseado no \textit{Concern Diffusion over operations (CDO)}, e (iii) linhas de código baseado no \textit{Concern Diffusion over Lines of Code (CDLOC)} \cite{Figueiredo, Garcia}. Estas métricas ajudam a revelar o grau de espalhamento e entrelaçamento das funcionalidades (\textit{features}) implementadas nos módulos da aplicação alvo. Quanto menor o número de módulos necessários para a implementação de uma funcionalidade, menor será o seu grau de espalhamento. Quanto maior o número de funcionalidades existentes em um determinado módulo, maior será o grau de entrelaçamento entre elas. As métricas do SoC foram quantificadas manualmente~\cite{Figueiredo, Garcia} através do ``sombreamento'' manual do código que identifica quais trechos do código fonte contribuem para a implementação de uma determinada funcionalidade~\cite{Figueiredo}. 

\textbf{Acoplamento.} Esta variável busca quantificar, através de duas métricas Dep\_Out e Dep\_In, o quão os elementos de \textit{design} (pacotes, classes e métodos) estão acoplados. Quanto maior o grau de dependência entre eles, mais elementos tendem a sofrer com propagações indesejadas de modificações. Dep\_In quantifica o número de classes fora de um pacote que depende das classes dentro deste pacote. Dep\_Out quantifica o número de classes dentro de um pacote que dependem de classes fora deste pacote.

\textbf{Complexidade e coesão.}  A complexidade mede o grau de conectividade entre os elementos por pacote. Assim, para seu cálculo foram somados os valores dos pacotes do projeto. Não obstante, a coesão mensura o grau em que os elementos estão logicamente relacionados ou ``pertencem um ao outro''. Por consequência, quanto maior a conectividade entre os elementos maior a coesão. Assim como a complexidade, seus valores são por pacote, neste caso foram somados os valores e divididos pela quantidade de pacotes. Tanto que ambas métricas estão um tanto relacionadas ao tamanho.

\textbf{Tamanho.} As métricas referente ao tamanho são bastante objetivas, incluindo linhas de código (LOC), número de atributos (NumAttr) e operações (NumOps). Tanto que as linhas de código foram coletadas manualmente, desconsiderando linhas em branco ou comentários. Já o número de atributos e operações foi coletado em partes manualmente, e, em outras através do SDMetrics\footnote{SDMetrics: https://www.sdmetrics.com/}. Todavia, para todas métricas referente ao tamanho, foram contabilizadas apenas aquelas que contribuíram para o cenário de evolução que estava sendo avaliado~\ref{subsec:cenarios-de-mudanca}.

\subsection{Processo experimental}
\label{sec:metodologia-processo}

\begin{figure}[ht!]
    \centering       
    %\begin{minipage}{.5\textwidth}
    \includegraphics[scale=.27]{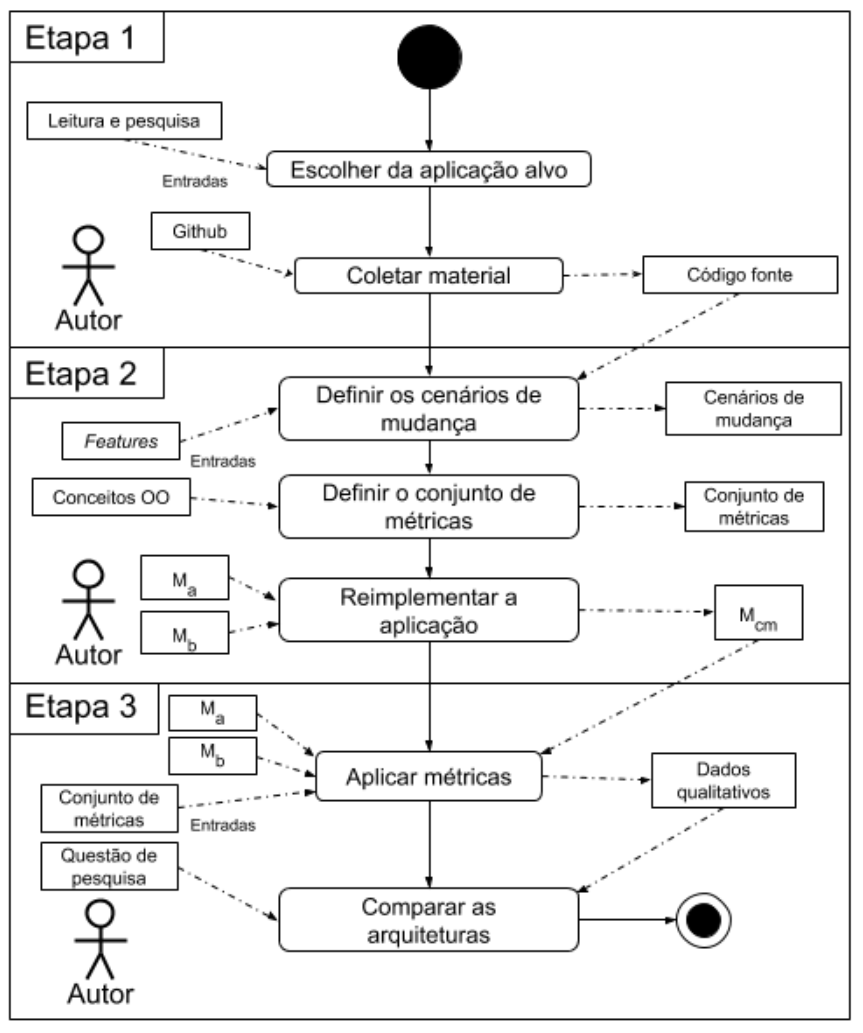}
		% \fonte{\cite{computacaoUnisinos}}
    %\end{minipage}
    \caption{Processo experimental}
    \label{fig:processo-experimental}
\end{figure}

Figura~\ref{fig:processo-experimental} apresenta o processo experimental adotado, o qual é formado por três etapas, incluindo a (1) identificação de aplicação alvo, (2) implementação e coleta de dados e (3) análise dos resultados. Cada etapa é discutida a seguir.

\textbf{Etapa 1: Identificação de aplicação alvo.} A primeira etapa se concentrou em buscar na indústria, uma aplicação alvo realística que utilizasse arquitetura dirigida a eventos. Neste sentido, identificou-se a aplicação desenvolvida por Ben Stopford~\cite{stopford2018designing, Ben} como sendo a aplicação alvo (descrita em Seção~\ref{sec:metodologia-aplicacao}). Esta aplicação foi desenvolvida utilizando boas práticas, trata-se de uma aplicação real que utiliza tecnologias amplamente utilizadas pela a indústria, tais como o Apache Kafka.

\textbf{Etapa 2: Implementação e coleta de dados.} As funcionalidades da aplicação alvo foram identificadas e organizadas em cenários de evolução, de tal forma que permitissem a implementação de uma aplicação similar utilizando o estilo arquitetural REST. Destaca-se que as funcionalidades da aplicação alvo estão bem documentadas em~\cite{stopford2018designing, Ben}. Após a identificação, os funcionalidades foram implementadas utilizando o \textit{framework Spring Boot} e~\textit{Spring Web}. Inevitavelmente, por se tratar de arquiteturas distintas, certas diferenças e refatorações são esperadas, a fim de alinhar as aplicações com suas evoluções. Após a implementação de cada cenário de mudança, será coletado um conjunto de métricas definido na Seção~\ref{sec:metodologia-metricas}, utilizando a ferramenta \textit{SDMetrics} para coletar parte das métricas. A segunda etapa foi finalizada após a implementação da aplicação alvo utilizando o estilo arquitetural REST.

\textbf{Etapa 3: Análise dos resultados.} Conforme citado na etapa anterior, a partir das métricas obtidas pela ferramenta \textit{SDMetrics}, foi possível realizar comparações entre as aplicações --- as quais serão essencialmente entre as aplicações após cada implementação, permitindo observar suas evoluções. Entretanto, outras comparações podem ser realizadas como, por exemplo, a comparação entre as versões da mesma aplicação, a fim de analisar sua evolução. Assim, para auxilio da análise, serão utilizadas tabelas que permitem identificar visualmente mudanças, além de classificar os elementos por uma métrica e destacar os elementos por percentuais.

\subsection{Procedimento da análise}
\label{sec:metodologia-procedimento}

Gráficos de linha são usados para fornecer uma visão geral dos dados coletados no processo de medição. Esses gráficos nos permitem analisar o impacto da arquitetura dirigida a eventos nas métricas de modularização definidas. Cada gráfico enfoca os dados coletadas em relação à uma métrica específica. O eixo X especifica os cenários de evolução. O eixo Y apresenta os valores coletados para uma métrica específica. Para trazer uma análise da distribuição dos dados, métodos estatísticos foram utilizados, incluindo desvio padrão, mediana e a média. Além disso, a diferença entre as médias foi contabilizada.

A análise quantitativa dos dados será através das métricas coletadas pela ferramenta SDMetrics, o qual contabiliza automaticamente as métricas Dep\_Out, Dep\_In, H, R, NumAttr e NumOps. As métricas de separação de interesses (CDC, CDO e CDLOC) foram contabilizadas de forma manual.

\subsection{Aplicação alvo}
\label{sec:metodologia-aplicacao}

\begin{figure}[ht!]
	
	\centering%
	%\begin{minipage}{.9\textwidth}
	\includegraphics[scale=0.3]{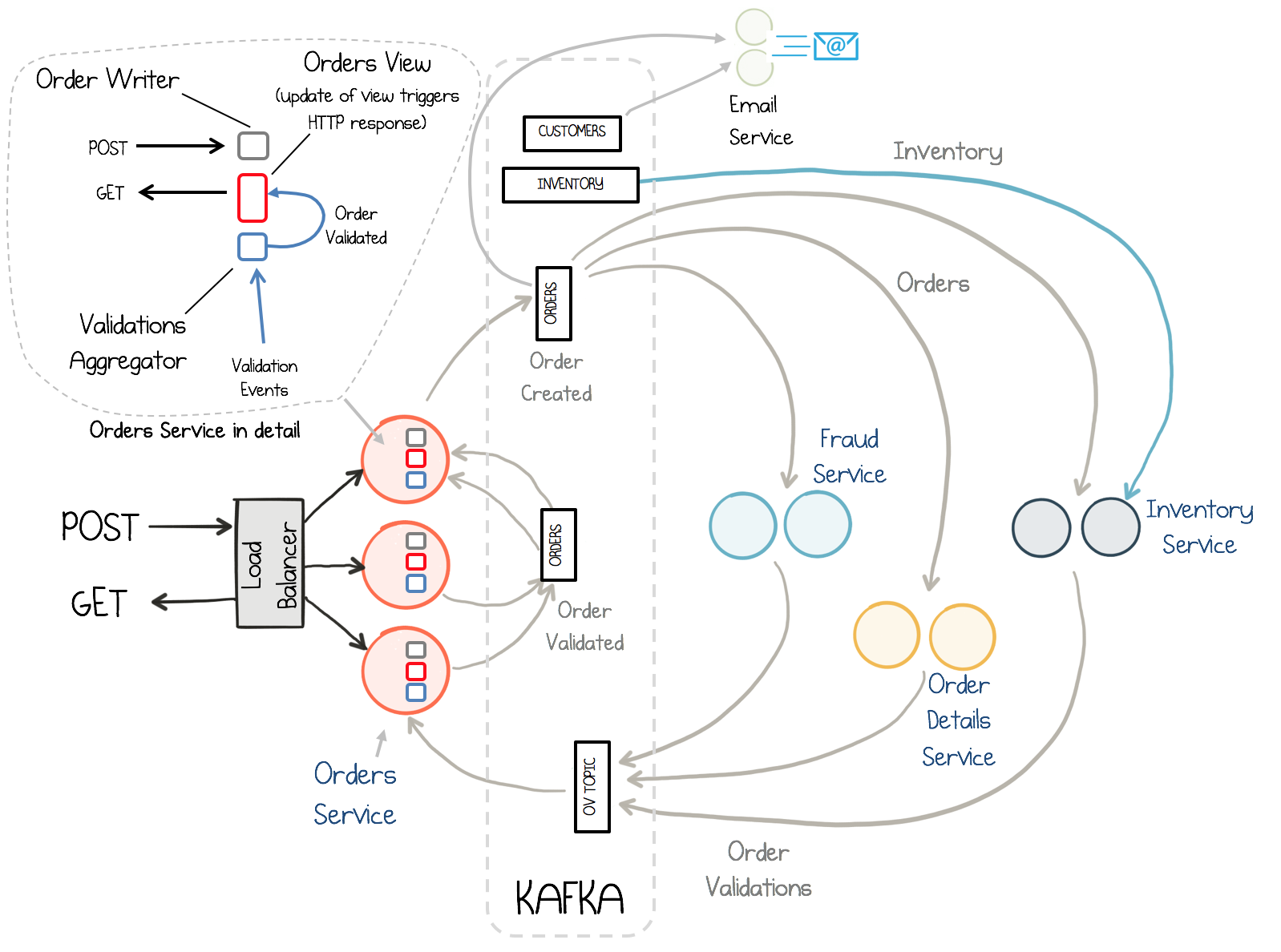}
        \caption{Ilustração esquemática da aplicação utilizada (fonte~\cite{Ben}).}
        \label{fig:diagram-application}
	%\fonte{\cite{Ben}}
	%\end{minipage}
\end{figure}

Figura~\ref{fig:diagram-application} apresenta uma ilustração esquemática da aplicação alvo. Trata-se um sistema de gerenciamento de pedidos composto por vários componentes. A aplicação alvo utilizada foi obtida a partir de um exemplo disponibilizado pela \textit{Confluent} em \cite{Ben}. Sendo assim, as principais razões para sua escolha foram o cuidadoso detalhamento da aplicação em \cite{Ben}, a disponibilidade da aplicação e a adoção de boas práticas de implementação. Pode-se considerar uma aplicação complexa, devido os recursos utilizados. A arquitetura abordada é a dirigida a eventos em conjunto com a arquitetura de microsserviços, em uma implementação utilizada a linguagem Java. O Kafka é o \textit{middleware} responsável por gerenciar a aplicação, como armazenamento dos dados, mapeamento de entidades, produção e consumo dos eventos. No site da Confluent, detalhes dos recursos utilizados e responsabilidades de cada componente são apresentados, além de disponibilizar o código fonte da implementação no repositório do \textit{Github} \footnote{https://github.com/confluentinc/kafka-streams-examples/tree/6.0.0-post/src/main/java/io/confluent/examples/streams/microservices}.

\subsection{Cenários de evolução}
\label{subsec:cenarios-de-mudanca}

Tabela~\ref{tab:tabela-cenarios} apresenta os cenários de evolução considerados. No total, cinco cenários foram identificados, cada qual contendo funcionalidades relativas à aplicação alvo. Cada cenário incorpora uma funcionalidade na versão anterior. O serviço de pedidos representa o ponto de entrada, uma interface REST provê os métodos \textit{POST} e \textit{GET} \cite{Ben}. Ao realizar um \textit{POST}, criará um evento no Kafka, que será consumido por outros três serviços de validação (validação do pedido,  identificação de fraude e reserva de estoque). Sendo que o pedido será validado em paralelo, emitindo \textit{PASS} (válido) ou \textit{FAIL} (inválido) baseado no sucesso de cada serviço \cite{Ben}. Como as validações ocorrem em paralelo, o resultado de cada uma é enviado por meio de seu tópico, separado das demais \cite{Ben}. Por fim, os resultados são agregados no serviço de pedidos aonde os pedidos são movidos para o status de \textit{PASS} ou \textit{FAIL}, baseado na combinação de resultados \cite{Ben}.

Para que o usuário consulte algum pedido a partir do \textit{GET}, no serviço de pedidos foi criado uma \textit{view} materializada consultável (``Orders view'' na Figura~\ref{fig:diagram-application}), usando um armazenamento de estado em cada instância do serviço, de forma que qualquer pedido pode ser requisitado historicamente \cite{Ben}. Dentre as validações, têm-se três: (i) validação do pedido confere os elementos básicos, como quantidade e preço do próprio pedido; (ii) identificação de fraude rastreia o valor total dos pedidos de cada cliente em uma janela de uma hora, alertando se o limite que configura uma fraude for alcançado; e (iii) reserva de estoque verifica se há unidades disponíveis para esse pedido \cite{Ben}, primeiramente precisa consultar quantas unidades disponíveis há em estoque, após, a quantidade requisitada será reservada pelo tempo necessário até que o pagamento seja concluído.

\begin{table}[ht!]
    \small
    \centering
    \caption{Descrição dos cenários de mudança}
    \begin{tabular}{|c|l|}
        \hline
        \textbf{Versão} & \textbf{Descrição} \\ \hline
        V1      & Serviço de pedidos \\ \hline
        V2      & Serviço de validação do pedido \\ \hline
        V3      & Serviço de identificação de fraude \\ \hline 
        V4      & Serviço de reserva de estoque \\ \hline
        V5      & Serviço de envio de \textit{e-mail} ao cliente \\ \hline
    \end{tabular}
    \label{tab:tabela-cenarios}
    %\fonte{Elaborado pelo autor.}
\end{table}

\begin{table*}
    \small
    \centering
    \caption{Resultados obtidos}
    \begin{tabular}{|c|l|l|l|l|l|c|} 
        \hline
        \textbf{Atributos}  & \textbf{Métricas}  & \multicolumn{1}{c|}{\textbf{Arquitetura}} & \multicolumn{1}{c|}{\begin{tabular}[c]{@{}c@{}}\textbf{Desvio} \\\textbf{padrão}\end{tabular}} & \multicolumn{1}{c|}{\textbf{Mediana}} & \multicolumn{1}{c|}{\textbf{Média}} & \textbf{Diferença} \\ 
    
        \hline
    
        \begin{tabular}[c]{@{}c@{}}SoC \\ \end{tabular}
        & \begin{tabular}[c]{@{}c@{}}CDC \\\end{tabular}
           & KAFKA & 1.72 & 3 & 3.2 &  40.74\%  \\ \cline{3-6}
          & & REST  & 2.50 & 4 & 5.4 &                           \\ 
        \cline{2-7}
        & \begin{tabular}[c]{@{}c@{}}CDO \\ \end{tabular}
            & KAFKA & 13.11 & 10 & 15.2 & 22.37\%  \\ \cline{3-6}
          & & REST  & 9.02  & 10 & 11.8 & \\ 
        \cline{2-7}
        & \begin{tabular}[c]{@{}c@{}}CDLOC \\ \end{tabular}         
            & KAFKA & 2.19 & 4 & 4 & 48.72\%  \\ \cline{3-6}
          & & REST  & 2.56 & 8 & 7.8 & \\ 
        
        \hline
    
        Coupling
        & \begin{tabular}[c]{@{}c@{}}Dep\_Out \\ \end{tabular}     
            & KAFKA & 3.83 & 4 & 5.4 & 70.37\%  \\ \cline{3-6}
          & & REST  & 0.80 & 1 & 1.6 & \\ \cline{2-7}
        & \begin{tabular}[c]{@{}c@{}} Dep\_In \\ \end{tabular}      
            & KAFKA & 3.83 & 4 & 5.4 & 85.19\%  \\ \cline{3-6}
          & & REST  & 0.75 & 1 & 0.8 & \\ 
        
        \hline
    
        Cohesion                                                        
        &  H 
           & KAFKA & 0.04 & 1.2075 & 1.207 & 20.94\%  \\ \cline{3-6}
         & & REST  & 0.02 & 0.966  & 0.9543 & \\ 
         
        \hline
    
        Complexity
        & R
            & KAFKA & 11.43 & 56 & 57.8 & 85.12\%  \\ \cline{3-6}
          & & REST  & 5.46  & 6  & 8.6  & \\ 
          
        \hline
    
        Size                                                             
        & LOC
           & KAFKA & 116.17 & 174 & 233 & 43.35\%  \\ \cline{3-6}
         & & REST  & 93.11  & 100 & 132 & \\ \cline{2-7}
        & \begin{tabular}[c]{@{}c@{}} NumAttr \\ \end{tabular}            
           & KAFKA & 4.83 & 8 & 9.2 & 41.30\%  \\ \cline{3-6}
         & & REST & 3.50  & 4 & 5.4 & \\ \cline{2-7}
        & \begin{tabular}[c]{@{}c@{}} NumOps \\ \end{tabular} 
            & KAFKA & 13.50 & 10 & 15.4 & 12.99\%  \\ \cline{3-6}
          & & REST  & 8.73  & 11 & 13.4 & \\
          
        \hline
    \end{tabular}
    \label{tab:results}
\end{table*}

\section{Resultados}
\label{sec:resultados}

Esta seção apresenta os resultados coletados após a execução da metodologia definida na Seção~\ref{sec:metodologia}. A Figura~\ref{fig:resultsSeparation}, Figura~\ref{fig:resultsCoupling} e Figura~\ref{fig:resultsSize} apresentam os resultados obtidos nos cenários de evolução. A Tabela~\ref{tab:results} traz indicadores estatísticos sobre os resultados, incluindo o desvio padrão, mediana, média e a diferença entre as médias.

\subsection{Separação de Interesses e Acoplamento} 

\begin{figure*}[ht!]
    \centering      
    %\begin{minipage}{\textwidth}
     \includegraphics[scale=1.3]{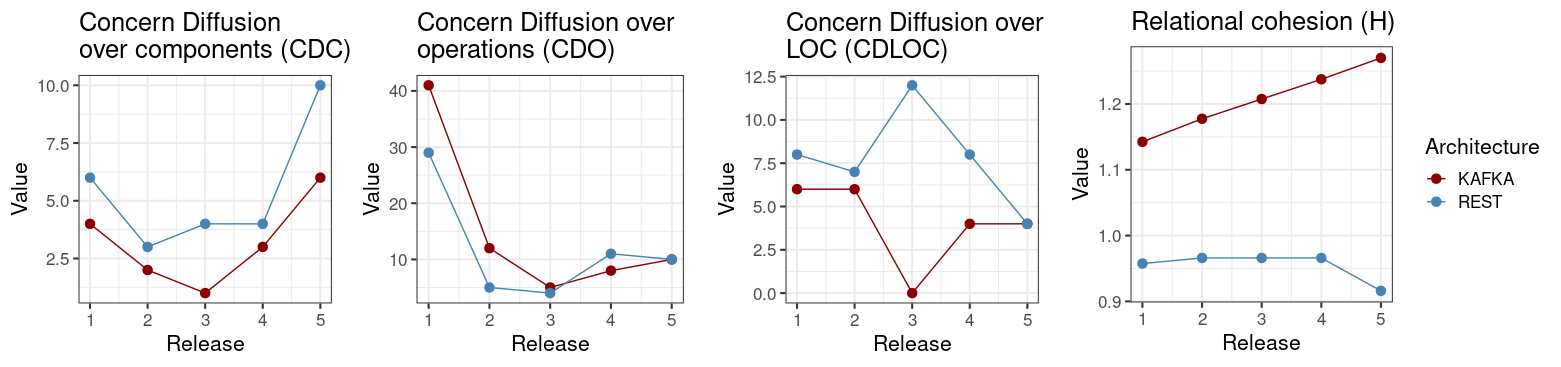}	
        %\fonte{\cite{Ben}}
    %\end{minipage}
    \caption{Resultados coletados considerando as métricas CDC, CDO, CDLOC e H.}
    \label{fig:resultsSeparation}
\end{figure*}

\begin{figure*}[ht!]
    \centering    
    %\begin{minipage}{\textwidth}
    \includegraphics[scale=1.4]{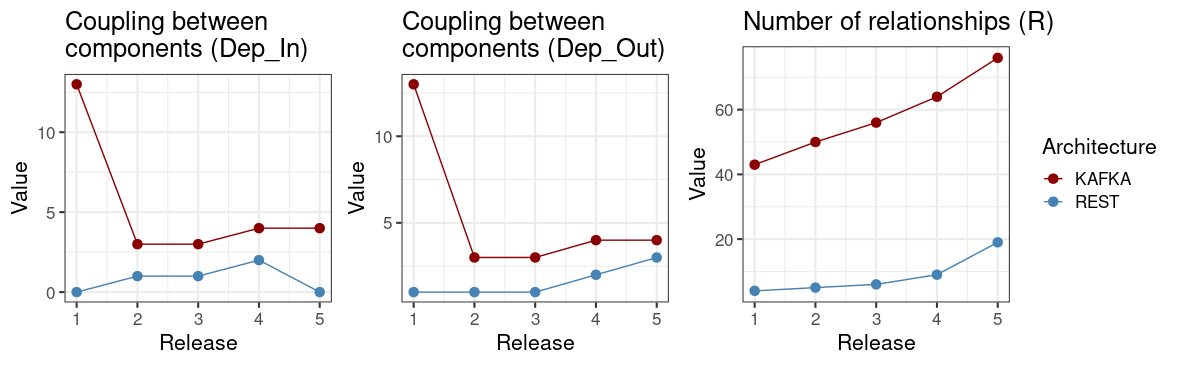}
        %\fonte{\cite{Ben}}
    %\end{minipage}
    \caption{Resultados coletados considerando as métricas Dep\_In, Dep\_Out e R.}
    \label{fig:resultsCoupling}
\end{figure*}

A Tabela~\ref{tab:results} mostra os resultados dos efeitos da arquitetura-dirigida a eventos na separação de interesses através da perspectiva de três métricas: CDC, CDO e CDLOC. O KAFKA apresentou resultados menores em comparação à arquitetura REST, considerando as métricas CDC e CDLOC. Isso pode ser notado através das diferenças entre as médias computadas 40,74\% e 48,72\%, respectivamente. Esse resultado indicada que menos classes (CDC) são afetadas em cada cenário de evolução, já que os serviços são independentes, mas compartilham classes auxiliares (\textit{utils}). Também observam-se números menores na quantidade de transições de interesses sob as linhas (CDLOC). Isso significa que o Kafka promoveu uma melhor modularização de \textit{concerns} considerando componentes e linhas de código. Na Figura~\ref{fig:resultsSeparation}, observa-se que em todas as versões, ambas métricas apresentaram valores menores para o KAFKA. Porém, o KAFKA apresentou resultados superiores para a métrica CDO, tendo uma diferença entre as médias de 22,37\%. A mediana da CDO, por sua vez, não registrou diferença. Na maioria dos cenários foram necessárias mais operações (ou funções) para implementar o serviço de cada cenário na arquitetura dirigida a eventos. Na Figura~\ref{fig:resultsSeparation}, observa-se que apenas na versão 4, o REST apresentou um valor superior ao Kafka. Logo, essa necessidade foi refletida na métrica CDO.

Considerando as variáveis de acoplamento da Tabela~\ref{tab:results}, observa-se que a arquitetura REST entregou menor acoplamento do que o KAFKA. Ambas métricas Dep\_out e Dep\_in, apresentam diferenças entre as médias computadas de 70,37\% e 85,19\%, respectivamente. Contudo, ao analisar a Figura~\ref{fig:resultsCoupling} percebe-se valores bastante elevados para ambas métricas no KAFKA na primeira versão. Mesmo que nos demais cenários os valores sejam superiores, a diferença não é tão alta quanto as diferenças entre as médias. Alias, não só um menor acoplamento na arquitetura REST, mas também uma menor variação em cada cenário de evolução, conforme observado na Figura~\ref{fig:resultsCoupling}.

A maior separação de interesses está em linha com a característica de microsserviços \cite{Falatiuk, laigner2020monolithic}, na qual cada serviço é independente. Isso promove uma maior modularização da aplicação, a qual se beneficia em cenários onde há alterações de comportamento ou evolução da aplicação, como a adição de novas \textit{features} \cite{SANTANNA}. Além disso, também evita-se a degradação da qualidade da aplicação, garantindo ao usuário acesso ao serviço sem interrupções. Ou ainda, traz benefícios a performance do projeto \cite{Subramanian}. Contudo, a independência dos serviços tem seu preço, resultando em mais funções, atributos e classes auxiliares. O maior acoplamento na arquitetura dirigida a eventos pode ser explicado pela necessidade de classes auxiliares. Indispensáveis para não gerar duplicação de código, já que as funções atendem diferentes contextos, caso não existissem gerariam mais código em cada serviço, consequentemente complexidade \cite{SANTANNA}.

\begin{tcolorbox}[colback=green!3]
\textbf{Resultados observados 1:} 
\textit{As médias de separação de interesses do KAFKA, das métricas CDC e CDLOC apresentaram diferenças de 40,74\% e 48,72\%, respectivamente. Além de valores menores em todas as versões. Evidenciando a maior separação de interesses se comparado ao REST. Entretanto, as médias de acoplamento mostram diferenças acima dos 70\% para o KAFKA. Isso mostra que as dependências, tanto internas quanto externas são maiores no KAFKA.} %
\end{tcolorbox}

\subsection{Complexidade, Coesão e Tamanho}

\begin{figure*}[ht!]
    \centering    
    %\begin{minipage}{\textwidth}
    \includegraphics[scale=1.4]{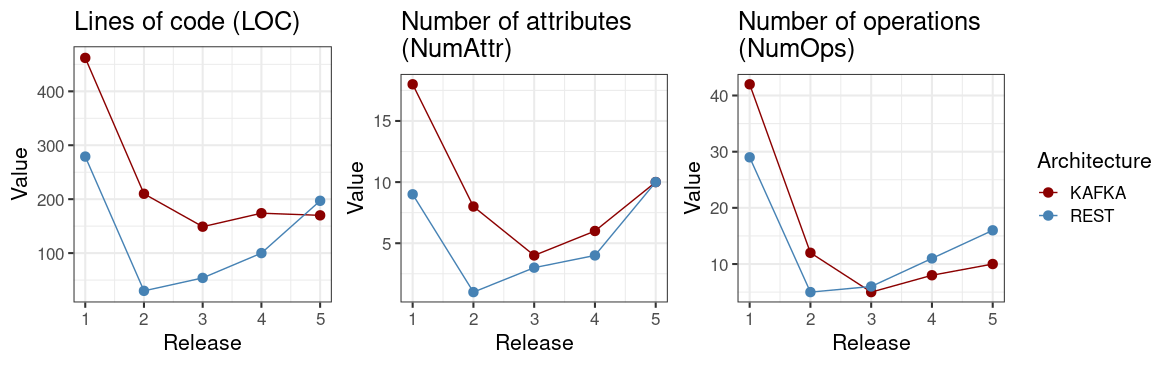}
        %\fonte{\cite{Ben}}
    %\end{minipage}
    \caption{Resultados coletados considerando as métricas LOC, NumAttr e NumOps.}
    \label{fig:resultsSize}
\end{figure*}

A Tabela~\ref{tab:results} traz os resultados dos efeitos da arquitetura-dirigida a eventos em relação à complexidade, coesão e tamanho ao longo dos cenários de evolução. Os resultados apontam que o KAFKA obteve resultados superiores em comparação à arquitetura REST, observando as variáveis de complexidade e coesão computadas através das métricas H e R, respectivamente. Os números obtidos apontam que as diferenças entre as médias computadas das métricas H e R foram 20,70\% e 85,12\%, respectivamente. Embora os resultados da coesão relacional (H) tenham favorecido ao Kafka, a complexidade deve ser um fator a ser avaliado.

Analisando os resultados através da perspectiva da variável tamanho, eles indicam que o KAFKA obteve resultados superiores em comparação à arquitetura REST. Esse achado foi quantificado utilizando as métricas LOC, NumAttr e NumOps, as diferenças entre as médias dessas métricas foram 43,34\%, 41,30\% e 12,99\% respectivamente. Por outro lado, analisando a Figura~\ref{fig:resultsSize}, observa-se valores bastante elevados nos três primeiros cenários para as métricas NumAttr e NumOps. Após, a métrica NumAttr apresenta valores elevados, mas bem próximos, e, a métrica NumOps a partir do terceiro cenário valores inferiores.

Tais resultados eram esperados após a implementação dos cenários, pois a quantidade de linhas e operações no KAFKA foram bastante elevados, principalmente nos primeiros cenários. Isso pode ser explicado pelo fato de que na arquitetura REST precisa-se de uma quantidade menor de elementos (classes) para a construção da aplicação. Por outro lado, no KAFKA observou-se a necessidade de criar várias classes auxiliares aos serviços. Bem como funções e atributos para configurações de cada serviço, tanto na sua inicialização como operação. Por consequência, essa necessidade refletiu em maiores valores na complexidade e coesão --- quais estão de alguma forma ligados ao aumento do tamanho da aplicação sem um gerenciamento adequado, por exemplo, entre o número de relacionamentos entre as classes e interfaces por pacote. Em suma, aplicações maiores tendem a também serem mais complexas, além de que outros estudos também perceberam maior complexidade na arquitetura dirigida a eventos \cite{Falatiuk, laigner2020monolithic}.

\begin{tcolorbox}[colback=green!3]
\textbf{Resultados observados 2:} 
\textit{As médias de complexidade e tamanho favoreceram o REST, destacando-se as métricas LOC e NumAttr com diferenças de 43,45\% e 41,30\%, respectivamente. Oriundos em sua maior fração de classes auxiliares, e, funções/atributos necessários ao funcionamento de cada serviço do KAFKA. Diretamente ligado a complexidade, medido pela métrica R que apresentou uma diferença de 85,12\% a favor do REST. Por outro lado, a média de coesão (H), favoreceu o KAFKA em 20,94\%.} %
\end{tcolorbox}

\subsection{Discussão}
\label{subsec:discussion}

A arquitetura dirigida a eventos obteve bons resultados na separação de interesses. Por outro lado, ao analisar os resultados obtidos pelo conjunto de métricas definidos, percebeu-se que a arquitetura REST obteve melhores resultados de uma maneira geral. Uma melhor separação de interesses viabiliza uma melhor modularização das funcionalidades, algo altamente recomendável quando busca-se ter serviços independentes e que evitem concatenação de modificações ao serem alterados \cite{SANTANNA}. A seguir, são abordadas algumas das características da arquitetura dirigida a eventos, por consequência do KAFKA também.

\textbf{Livre de coordenação por design.} O controle das consistências fica comprometido, ao enviar dados para muitos serviços diferentes. Uma forma de mitigar isso é adotar o principio de \textit{single writer}, pois consiste em delegar a tarefa de propagar eventos de um tipo especifico, um único serviço \cite{Ben}. Tal controle é mais evidente, ao replicar os eventos entre replicas (ou instâncias), algo normal e comum por questão de garantia. Assim, ao centralizar no \textit{single writer} criasse um ``túnel'' de consistências, validações e outros escritores por meio de um único fluxo. Tanto que, uma das características da arquitetura é mover os dados (ou eventos) enquanto opera neles \cite{Ludger}, no KAFKA através do \textit{Kafka Streams} e KSQL --- pontos centrais para processar dados dentro de programas clientes.

Eventos também são uma ferramenta útil ao \textit{design} do sistema, fornecendo notificação, transferência de estado e desacoplamento \cite{Ludger}. Embora o KAFKA apresentou maior acoplamento, segundo as métricas Dep\_in e Dep\_out da Figura~\ref{fig:resultsSeparation}. Visto que, há várias dependências por serviços auxiliares, comuns a todos os serviços. Contudo, no desenvolvimento dos cenários as dependências se limitaram a classes auxiliares, e, não aos outros serviços. Inclusive, no Euphoria \cite{schipor2019euphoria} foi identificado o desacoplamento, pois mudanças em um módulo não determinou mudanças em módulos adjacentes, desde que a interface permanecesse a mesma. Portanto, apesar das métricas reportarem maior acoplamento, não observou-se a concatenação de modificações nos serviços.

\textbf{Source of truth.} Caso os eventos sejam armazenados na ordem de criação e não sofrerem alterações. O \textit{log} fornece uma linha de tempo do que exatamente aconteceu \cite{stopford2018designing}. Logo, esse comportamento torna o fluxo de eventos a \textit{source of truth}, já em sistemas tradicionais é o banco de dados. Assim como, a aplicação explorada consiste em vários serviços cooperando em um único fluxo de negócio, onde cada um faz seu trabalho de processar eventos e criar novos --- eventos são o ponto central do sistema \cite{stopford2018designing}. Pois, o \textit{log} disponibiliza os dados centralmente como a \textit{shared source of truth} aos serviços, porém com um contrato simples, o que mantém os serviços fracamente acoplados. Desta forma, o CQRS é um dos pontos chave, pois ele separa o caminho de gravação do caminho de leitura e os vincula a um canal assíncrono. Seu funcionamento é no estilo \textit{log write-ahead}, as inserções e atualizações são imediatamento registradas sequencialmente no disco. Desse modo, torna o processo bastante rápido pois não precisa esperar pelo lento processo de atualizar várias estruturas como tabelas, índices e assim por diante \cite{stopford2018designing}.

A garantia do devido funcionamento da aplicação, requer alguns cuidados. Primeiro, a ordem dos eventos deve ser preservada, ao serem enviados sempre a mesma partição isso garante a correta ordenação. Já para garantias de ordem global, deve-se utilizar um tópico de partição única, isso vai limitar a taxa de transferência, se bem que é o suficiente para a maioria dos casos \cite{stopford2018designing}. Segundo, o reenvio de eventos, causado por alguma falha de rede, \textit{garbage-colletor} demorado, falha ou algo semelhante. Como as mensagens são enviadas em lotes, deve-se ter cuidado de enviar os lotes um por vez, por máquina de destino, a fim de evitar a reordenação dos eventos \cite{stopford2018designing}. Em vista disso, o Kafka fornece poderosas funcionalidades, mas que precisam ser devidamente configuradas, e, cuidados no desenvolvimento dos serviços a fim de evitar efeitos indesejáveis e difíceis de rastrear. Inclusive, em \cite{laigner2020monolithic} percebeu-se como desvantagem, o complexo fluxo de eventos acarretado pelo número de microsserviços. Portanto, a arquitetura dirigida a eventos apresenta maior complexidade, assim, seu desenvolvido e manutenções podem vir a requerer maior esforço e devido gerenciamento conforme evolução.

\textbf{Stream processing.} Há bastante tempo os sistemas de mensagens são utilizados para trocar eventos entre sistemas, porém apenas recentemente eles começaram a serem utilizados na camada de armazenamento. Isso cria um estilo arquitetônico interessante. Ao analisar a estrutura de um banco de dados, encontram-se uma série de componentes, análogo a uma caixa preta. Sendo assim, essa estrutura pode ser decomposta utilizando processamento de fluxo (\textit{streaming}) e esses componentes existirem em locais diferentes, unidos pelo \textit{log} \cite{stopford2018designing}. Logo, surgiram plataformas de \textit{streaming}, como o KAFKA que processa o fluxo de eventos, armazenar os eventos em estrutura de \textit{log} e disparar uma cascata de serviços inscritos a tópicos. Isso permite aplicativos e serviços incorporem lógica diretamente sobre o fluxos de eventos. Além de disponibilizar os recursos de processamento do banco de dados na camada de aplicação, por meio de uma API.

O \textit{Kafka Streams} é a principal API para processamento de \textit{stream} na JVM. Baseado em uma DSL (\textit{domain-specific language}), que fornece uma interface estilo declarativo onde os fluxos podem ser unidos, filtrados, agrupados ou agregados \cite{stopford2018designing}. Bem como, fornece mecanismos de estilo funcional como \textit{map}, \textit{flatMap}, \textit{transform}, \textit{peek}, entre outros \cite{stopford2018designing}. Diante disso, podem ser criados índices ou visualizações, e essas visualizações se comportam como uma forma de cache continuamente atualizado, dentro ou perto de seu aplicativo. A aplicação utilizada faz uso do \textit{Kafka Streams}. Como resultado, em geral as consultas e processamento de eventos na DSL, requerem mais linhas de código, quantificado na métrica LOC da Figura~\ref{fig:resultsSize}. Ao comparar a DSL com \textit{querys} SQL, percebe-se que a DSL é mais complexa, entretanto, mais poderosa também. Portanto, o uso do \textit{Kafka Streams} gera mais linhas de código, enquanto também entrega mecanismos mais poderosos.

\textbf{Escalabilidade.} Entre as características da arquitetura dirigida a eventos, a altamente escalabilidade demonstra ser a mais desejada \cite{schipor2019euphoria}. Sendo capaz de suportar diferentes tamanhos de \textit{clusters}, por exemplo 5, 100, 200 ou quantas máquinas forem necessárias \cite{stopford2018designing}. A adição de máquinas se resume em adicionar novas maquinas e rebalancear. Entretanto, podem aparecer problemas de negação de serviços, administráveis com o \textit{cotas}, um serviço para definir a largura de banda a cada serviço. Sua grande vantagem está no seu armazenamento (estrutura de \textit{logs}), pois não utilizam índices que são bastante custosos para se manter. Portanto, sua complexidade é O(1) ao ler e gravar mensagem em uma partição. Além de que, ao ler ou escrever em um tópico é capaz de replicar os eventos para todas instancias definidas. Suas consultas, abordadas anteriormente podem ser em paralelo, dividindo as mensagens de um tópico entre diversos consumidores.

Serviços orientados a eventos devem sempre ser executados com \textit{highly  available (HA)}, a menos que não haja requisitos para HA. A principal razão para isso é essencialmente um ambiente autônomo \cite{stopford2018designing}. Se há apenas uma instancia do serviço, ao adicionar uma segunda, a carga será rebalanceada naturalmente. O mesmo acontece caso um nó falhe. Portanto, os serviços herdam a alta disponibilidade e balanceamento de carga naturalmente, o que significa que eles podem escalar horizontalmente lidar com interrupções não planejadas ou realizar reinicializações continuas sem perder a operabilidade do serviço \cite{stopford2018designing}. Seu maior impacto pode ser observado na Figura~\ref{fig:resultsSize}, através da métrica LOC com valores consideravelmente superiores se comparado ao REST, e, NumOps que também apresentou valores superiores, mas com menor diferença. Isso, se dá pelas várias configurações (LOC) e funções (NumOps) em cada serviço para conexão com a plataforma e configuração dos tópicos utilizados por ele. Tamanhos maiores são um dos fatores que contribuem com a complexidade da aplicação, refletido na métrica R da Figura~\ref{fig:resultsCoupling}, também observados em \cite{laigner2020monolithic}. Portanto, a complexidade da arquitetura dirigida a eventos pode ser verificada em diferentes perspectivas.

\subsection{Limitações}
\label{subsec:limitation}

O estudo exploratório reportado trata-se de um estudo inicial que explora uma área ainda pouco investigada na literatura. Neste sentido, o estudo possui algumas limitações que precisam ser consideradas. Apenas uma aplicação foi considerada no estudo. Priorizou-se o uso de aplicações da indústria e de código aberto que fizessem uso da arquitetura dirigida a eventos. Outras aplicações foram encontradas, além da aplicação alvo, porém não foram consideradas devido à algumas restrições, tais como tamanho pequeno, não eram de código aberto, não adotaram boas práticas de implementação, e não possuem documentação. Assim como argumentado no estudo, aplicações dirigidas a eventos mostram uma complexidade elevada, que também está presente no desenvolvimento, pela sua curva de aprendizado percebida. Sendo esse um dos motivos pelo qual o estudo explora apenas uma aplicação, desenvolver uma além de inviável poderia prejudicar a avaliação dos resultados. Tal dificuldade, pode ser compreendida pela grande diferença se comparado a arquiteturas convencionais. Conforme explorado anteriormente, não há banco de dados, eventos são gatilhos para serviços e deve-se construir um fluxo de eventos.

\section{Conclusão e Trabalhos Futuros}
\label{sec:conclusion}

Arquitetura dirigida a eventos vem sendo adotada na indústria e algumas tecnologias foram propostas para viabilizá-las, tais como o KAFKA. Ela surge como uma alternativa à arquitetura REST para a implementação de sistemas orientados a serviços. Este trabalho reportou um estudo empírico inicial com o propósito de comparar a arquitetura dirigida a eventos e a arquitetura REST --- através da perspectiva de separação de interesses, acoplamento, coesão, complexidade e tamanho --- ao longo de 5 cenários de evolução de uma aplicação real.

A arquitetura dirigida a eventos, representada pelo KAFKA, apresentou bons resultados quanto à separação de interesses. Por outro lado, as demais métricas não apresentaram resultados melhores do que a arquitetura REST tradicional. Ao proporcionar uma melhor modularidade dos interesses, algumas características podem ser afetadas. Portanto, ao adotar uma arquitetura dirigida a eventos é preciso analisar tais vantagens e desvantagens apontadas neste estudo. Como trabalhos futuros, pretende-se realizar: (1) aumentar o número de métricas contabilizadas, visando aumentar a perspectiva de análise; (2) considerar mais aplicações para replicar o estudo realizado; e (3) coletar mais dados apra viabilizar uma análise estatística rigorosa. Esse trabalho pode ser visto como um primeiro passo para uma agenda mais robusta de estudos experimentais relacionados aos efeitos de arquiteturas dirigidas a eventos na modularidade de software.

%%
%% The next two lines define the bibliography style to be used, and
%% the bibliography file.
\bibliographystyle{ACM-Reference-Format}
\bibliography{sample-sigconf}

\end{document}